\documentstyle[preprint,aps,epsfig]{revtex}
\tightenlines
\begin{document}
\draft
%%%%%%%%%%%% Begin Cover Page %%%%%%%%%%%%%%%%%%%%%%%%%%%%%%%%%%%

\title{Drell-Yan Massive Lepton-Pair's Angular Distributions \\
       at Large $Q_T$} 
\author{Jianwei Qiu, Ricardo Rodriguez, and Xiaofei Zhang}
\address{
     Department of Physics and Astronomy,
     Iowa State University \\
     Ames, Iowa 50011, USA}

\date{February 14, 2001}
\maketitle
\begin{abstract}
By measuring Drell-Yan massive lepton-pair's angular distributions, we
can identify the polarization of the virtual photon of invariant mass
$Q$ which decays immediately into the lepton-pair.  In terms of a
modified QCD factorization formula for Drell-Yan process, which is
valid even if $Q_T\gg Q$, we calculate the massive lepton-pair's
angular distributions at large $Q_T$.  We find that the virtual
photons produced at high $Q_T$ are more likely to be transversely
polarized.  We discuss the implications of this finding to the
J/$\psi$ mesons' polarization measured recently at Fermilab.  
\end{abstract}
\vspace{0.2in}


%%%%%%%%%%%%%%%%%%%%%%%%%%%%%%%%%%%%%%%%%%%%%%%%%%%%%%%%%%%%%%%%%%%%%%%%%

Recent data on J/$\psi$ polarization measured by CDF collaboration at
Fermilab Tevatron seem to be inconsistent with the predictions from
various models of J/$\psi$ production \cite{CDF-jpsi}.  The CDF data
shows that the J/$\psi$ mesons produced at $\sqrt{S}=1.8$~TeV become
more longitudinally polarized as the transverse momentum ($Q_T$)
increases \cite{CDF-jpsi}.  On the other hand, the various theoretical 
calculations predict the J/$\psi$ mesons to be more transversely
polarized at large $Q_T$ \cite{BKL-jpsi}.  

\begin{figure}
\begin{center}
\epsfig{figure=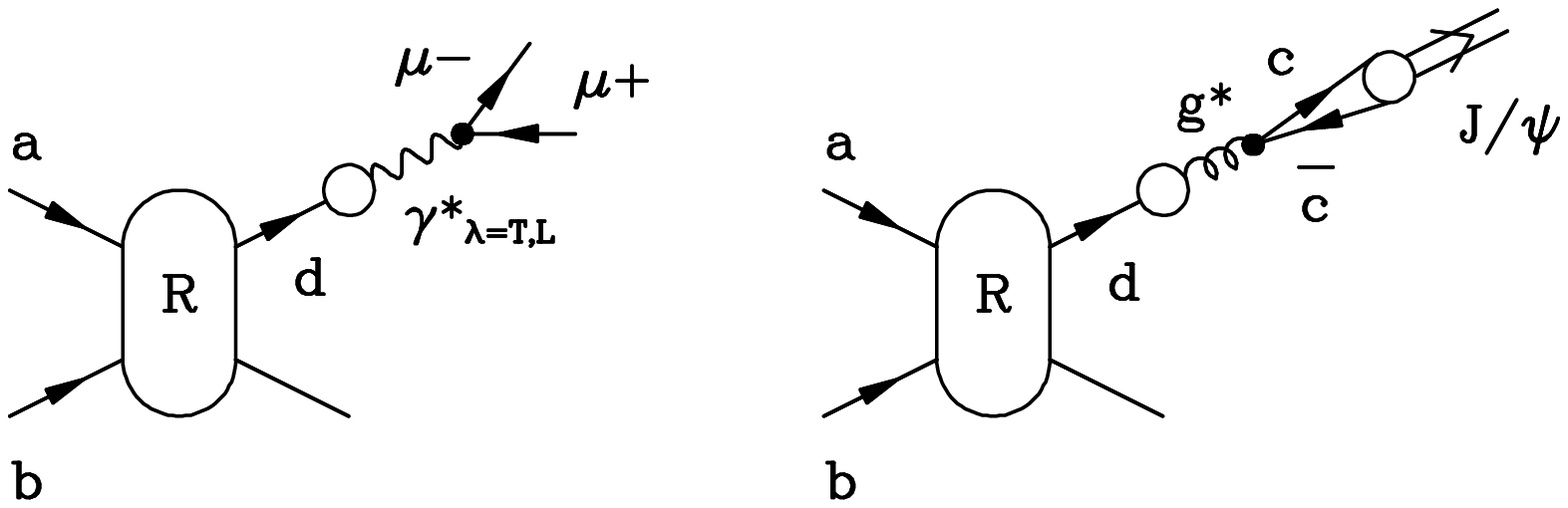,width=5.0in}
\end{center}
\caption{Sketch for Drell-Yan massive lepton-pair and J/$\psi$
production via parton fragmentation.}
\label{fig1}
\end{figure}

In order to help solving this puzzle, it was proposed to measure the
Drell-Yan massive lepton-pair's angular distributions at the
kinematics similar to that of J/$\psi$ production, and to extract the
virtual photon's polarization from the lepton-pair's angular
distributions \cite{QVZ-qm2001,QZ-vpff}.  Production of the
virtual photons in the Drell-Yan process at large $Q_T$ and small
$Q^2$ has a lot in common with the production of J/$\psi$ mesons at
high $Q_T$.  They both have two large 
physical scales: $Q_T$ and $Q^2$, which is equal to $M_{{\rm
J/}\psi}^2$ in the case of J/$\psi$ production; and $Q_T^2$ is much
larger than $Q^2$.  If the collision energy $\sqrt{S}$ is large
enough, and the logarithmic contributions from resumming all powers of
$\ln^n(Q_T^2/Q^2)$ dominate the production cross sections, the virtual
photon and J/$\psi$ production will share the {\it same} partonic
subprocesses, labeled by the $R$ in Fig.~\ref{fig1}.  Then, the only
difference between the virtual photon and J/$\psi$ production at large
$\sqrt{S}$ and high $Q_T$ is the difference in the fragmentation
functions. The virtual photon fragmentation functions are completely
perturbative if $Q^2\gg\Lambda_{\rm QCD}^2$ \cite{QZ-vpff,BL-DY},
while the parton to J/$\psi$ fragmentation functions involve
final-state nonperturbative soft interactions \cite{QS-jpsi}.  The
measurements of the virtual photon and J/$\psi$ polarization at high
$Q_T$ can help us to isolate the final-state effect in J/$\psi$
production, and to narrow the questions about J/$\psi$ formation.

In this letter, we calculate Drell-Yan massive lepton-pair's angular
distributions in Quantum Chromodynamics (QCD).  We present the
polarization parameter $\alpha_{\rm DY}$ as a function of $Q_T$, which
can be in principle measured at 
Fermilab Tevatron.  We find that the virtual photons at high $Q_T$ are
likely to be transversely polarized, which is different from the data
on the polarization parameter $\alpha_{{\rm J/}\psi}$ measured at
Fermilab \cite{CDF-jpsi}.  If our predictions for the 
$\alpha_{\rm DY}$ are confirmed by future experiments, we may conclude
that the non-perturbative final-state interactions are very important
for the formation of J/$\psi$ mesons.  

%%%%%%%%%%%%%%%%%%%%%%%%%%%%%%%%%%%%%%%%%%%%%%%%%%%%%%%%%%%%%%%%%%%%%%%%%

When $Q_T$ and $Q^2$ are both large, cross sections for Drell-Yan
massive lepton-pair production in hadronic collisions, 
$A(P_A)+B(P_B)\rightarrow \gamma^*(\rightarrow l\bar{l}(Q))+X$, 
can be systematically calculated according to QCD factorization
theorem \cite{CSS-fac,GTB-fac}.  Since we are mainly interested in the 
cross sections at high $Q_T$ and low $Q^2$, we neglect the $Z$ channel
contributions in the following discussions.   

\begin{figure}
\begin{center}
\epsfig{figure=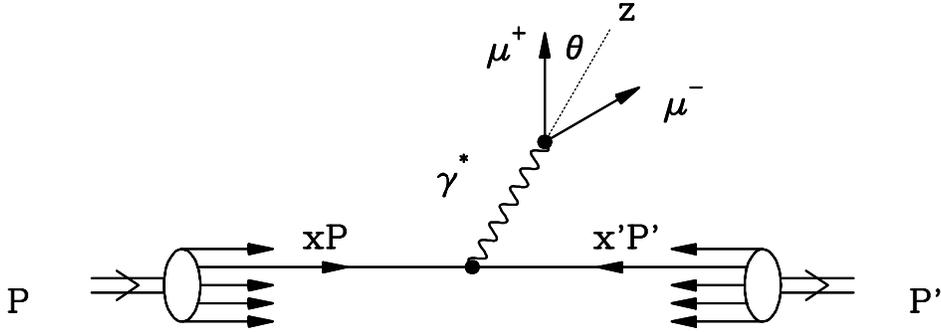,width=5.0in}
\end{center}
\caption{Definition of the angle $\theta$ in 
Eq.~(\protect\ref{DY-theta})
for Drell-Yan massive lepton-pair production.}
\label{fig2}
\end{figure}

After integrating over the azimuthal angle distribution, the
Drell-Yan lepton-pair's angular distributions, as sketched in 
Fig.~\ref{fig2}, can be expressed as \cite{LT-DY}
\begin{equation}
\frac{d\sigma_{AB\rightarrow l\bar{l}(Q)+X}}
     {dQ^2\, dQ_T^2\, dy\, d\cos\theta}
= \frac{d\sigma_{AB\rightarrow l\bar{l}(Q)+X}}
       {dQ^2\, dQ_T^2\, dy}\
  \frac{3}{2(3+\alpha_{\rm DY})}\, 
  \left[1+ \alpha_{\rm DY} \cos^2\theta \right]
\label{DY-theta}
\end{equation}
where the polarization parameter $\alpha_{\rm DY}$ is given in terms
of the cross sections for producing a polarized virtual photon, 
\begin{equation}
\alpha_{\rm DY} \equiv \left. 
\left(
 \frac{d\sigma_{AB\rightarrow \gamma^*_T(Q)+X}}{dQ_T^2 dy}
-\frac{d\sigma_{AB\rightarrow \gamma^*_L(Q)+X}}{dQ_T^2 dy}
\right) \right/
\left(
 \frac{d\sigma_{AB\rightarrow \gamma^*_T(Q)+X}}{dQ_T^2 dy}
+\frac{d\sigma_{AB\rightarrow \gamma^*_L(Q)+X}}{dQ_T^2 dy}
\right)\, .
\label{alpha}
\end{equation}
The subscripts, $T$ and $L$ in Eq.~(\ref{alpha}) represent the
transverse and longitudinal polarizations, respectively.  
The inclusive Drell-Yan cross section in Eq.~(\ref{DY-theta}) can also
be expressed in terms of the cross sections for producing a polarized
virtual photon,  
\begin{equation}
\frac{d\sigma_{AB\rightarrow l\bar{l}(Q)+X}}{dQ^2\,dQ_T^2\,dy}
= \left(\frac{\alpha_{em}}{3\pi Q^2}\right)
\left[2\, 
 \frac{d\sigma_{AB\rightarrow \gamma^*_T(Q)+X}}{dQ_T^2\, dy}
+\frac{d\sigma_{AB\rightarrow \gamma^*_L(Q)+X}}{dQ_T^2\, dy}
\right]
\label{DY-Vph}
\end{equation}
where the factor 2 is a consequence that the virtual photon has two
transverse polarization states \cite{LT-DY}.  

When $Q_T$ and $Q^2$ are both large, the differential cross section
for producing a virtual photon of polarization $\lambda=T$ or $L$ in
Eqs.~(\ref{alpha}) and (\ref{DY-Vph}) can be factorized as
\cite{CSS-fac,GTB-fac}  
\begin{equation}
\frac{d\sigma_{AB\rightarrow \gamma^*_\lambda(Q)+X}}{dQ_T^2\,dy}
= \sum_{a,b} 
    \int dx_1 \phi_{a/A}(x_1,\mu)\, 
    \int dx_2 \phi_{b/B}(x_2,\mu)\,
  \frac{d\hat{\sigma}_{ab\rightarrow\gamma^*_\lambda(Q)X}}
       {dQ_T^2\,dy}\left(x_1,x_2,Q,Q_T,y;\mu\right)
\label{Vph-fac}
\end{equation}
where $\sum_{a,b}$ run over all parton flavors, the $\mu$ represents
both renormalization and factorization scales, and
$\phi_{a/A}$ and $\phi_{b/B}$ are normal parton distributions. 
The short-distance parts, 
$d\hat{\sigma}_{ab\rightarrow\gamma^*_\lambda(Q)X}/dQ_T^2 dy$ in
Eq.~(\ref{Vph-fac}) are perturbatively calculable in terms of a power
series in $\alpha_s$ \cite{Arnold:1989dp}.  
When $Q_T\gg Q$, the perturbatively calculated short-distance parts
receive a large logarithm $\ln(Q_T^2/Q^2)$ for every power of
$\alpha_s$ beyond the leading order. Therefore, we need to resum such  
logarithms to all orders in $\alpha_s$ in order to derive a reliable
cross section for the virtual photon production at large $Q_T$.

It was shown \cite{BQZ-DY} that the large logarithms
$\ln^n(Q_T^2/Q^2)$ come from the partonic subprocesses in 
which the virtual photon is produced from the decay of a parton that
itself was produced at a distance scale $\sim 1/Q_T$.  Because the
parent parton can radiate massless partons before it radiates the
virtual photon, there is one power of large $\ln(Q_T^2/Q^2)$ for
an additional power of $\alpha_s$.  Since all powers of the large
logarithms are from the final-state bremsstrahlung of a parton
produced at a short-distance, such logarithms can be resumed into a
fragmentation function for the parton to fragment into the virtual
photon.  Berger, Qiu, and Zhang \cite{BQZ-DY} derived a modified
factorization formula for calculating Drell-Yan cross section at
$Q_T\ge Q$, which includes the resummation of the large logarithms,
\begin{eqnarray}
\frac{d\sigma_{AB\rightarrow \gamma^*_\lambda(Q)+X}}{dQ_T^2\,dy}
&=& \sum_{a,b,c} 
    \int \frac{dz}{z^2}\, 
    \int dx_1\, \phi_{a/A}(x_1,\mu)\, 
    \int dx_2\, \phi_{b/B}(x_2,\mu)
\nonumber \\
&\ & {\hskip 0.5in} \times
\left[
  \frac{d\hat{\sigma}_{ab\rightarrow cX}}{dp_{c_T}^2\,dy}
  \left(x_1,x_2,p_c=\hat{Q}/z;\mu\right) \right]\,
  D_{c\rightarrow \gamma^*_\lambda}(z,\mu;Q)
\nonumber \\
&+&
\sum_{a,b} 
    \int dx_1 \phi_{a/A}(x_1,\mu)\, 
    \int dx_2 \phi_{b/B}(x_2,\mu)\,
  \frac{d\hat{\sigma}_{ab\rightarrow\gamma^*_\lambda(Q)X}^{Y}}
       {dQ_T^2\,dy}\left(x_1,x_2,Q,Q_T,y;\mu\right)
\label{HQT-fac}
\end{eqnarray}
where $a,b,$ and $c$ run over all parton flavors, $\hat{Q}^\mu$ is the
on-shell part of momentum $Q^\mu$,  and $\mu$ represents
the renormalization, factorization, and fragmentation scales.  The
first term on the right-hand-side of Eq.~(\ref{HQT-fac}) represents
the resummed contributions, which include all powers of the large
logarithms; and the second term is free of the large logarithms and
carries the same physical meaning as the $Y$-term in the resummation
formula for $Q_T\ll Q$ \cite{Collins:1985kg}.  In Eq.~(\ref{HQT-fac}),
the $D_{c\rightarrow\gamma^*_\lambda}$ are the fragmentation functions
for the partons of flavor $c$ to fragment into a virtual photon of
invariant mass $Q$ and polarization $\lambda$, and their operator
definitions are given in 
Ref.~\cite{QZ-vpff}.  The resummation of the large $\ln(Q_T^2/Q^2)$ to 
all orders in $\alpha_s$ can be achieved by solving the evolution (or
renormalization group) equations for these fragmentation functions
\cite{QZ-vpff}.  In Eq.~(\ref{HQT-fac}), the partonic part 
$d\hat{\sigma}_{ab\rightarrow cX}/dp_{c_T}^2 dy$ are
perturbatively calculable and represent the production of partons of
flavor $c$ at a distance scale $\sim 1/p_{c_T}\sim 1/Q_T$
\cite{Owens:1987mp}.  Since the $Y$-term is defined to be the
perturbative difference between Drell-Yan cross section and the
resummed part of the cross section, the corresponding partonic parts
at the order of $\alpha_s^{n}$ are given by \cite{BQZ-DY}
\begin{equation}
\hat{\sigma}_{ab\rightarrow\gamma^*_\lambda(Q)X}^{Y(n)}
\equiv 
\hat{\sigma}_{ab\rightarrow\gamma^*_\lambda(Q)X}^{(n)}
- \sum_{m=2}^n 
\hat{\sigma}_{ab\rightarrow cX}^{(m)} \otimes 
D_{c\rightarrow\gamma^*_\lambda(Q)}^{(n-m)} 
\label{Y-hard}
\end{equation}
where $\otimes$ represents the convolution over $z$ as defined in
Eq.~(\ref{HQT-fac}), and $D^{(n)}$ are perturbatively calculated 
virtual photon fragmentation functions at order of $\alpha_s^{n}$. 

The modified factorization formula in Eq.~(\ref{HQT-fac}) should be
valid for all $Q_T\ge Q$.  When $Q_T\gg Q$, the resummed part
dominates the cross sections, and the $Y$-term represents a small
correction.  When $Q_T\sim Q$, the logarithms are small and the cross
sections are dominated by the $Y$-term.  All short-distance parts in  
Eq.~(\ref{HQT-fac}) are evaluated at the same short-distance scale
$\sim 1/Q_T$ and free of the large logarithms.  All leading
logarithmic contributions from a distance scale of $1/Q_T$ to $1/Q$
are resummed into the fragmentation functions.  With the modified
factorization formula in Eq.~(\ref{HQT-fac}), we are able to calculate
the polarization parameter $\alpha_{\rm DY}$ in Eq.~(\ref{alpha})
reliably in QCD perturbation theory by calculating the short-distance
parts order-by-order in $\alpha_s$. 

%%%%%%%%%%%%%%%%%%%%%%%%%%%%%%%%%%%%%%%%%%%%%%%%%%%%%%%%%%%%%%%%%%%%%%%%%

From the modified factorization formula in Eq.~(\ref{HQT-fac}), we
obtain the leading order contributions to the virtual photon cross
sections by calculating the lowest order contributions to both 
$\hat{\sigma}_{ab\rightarrow c}$ and
$\hat{\sigma}_{ab\rightarrow\gamma_\lambda^*}^{Y}$.
From Eq.~(\ref{HQT-fac}), the lowest order contributions to the
short-distance $d\hat{\sigma}_{ab\rightarrow cX}/dp_{c_T}^2 dy$ are
independent of the virtual photon's polarization, and are given by the
lowest order two-to-two partonic tree diagrams at $O(\alpha_s^2)$, 
\begin{equation}
\frac{d\hat{\sigma}_{ab\rightarrow cd}}{dp_{c_T}^2\,dy}
\left(x_1,x_2,p_c=\hat{Q}/z;\mu\right)
= \pi\, \frac{\alpha_s^2(\mu)}{x_1\,x_2\,S}
\left| \frac{1}{g^2}\overline{M}_{ab\rightarrow cd} \right|^2 \,
\delta(\hat{s}+\hat{t}+\hat{u})\, ,
\label{R-LO}
\end{equation}
where $g$ is the strong coupling constant and $S=(P_A+P_B)^2$ is the
collision energy. In Eq.~(\ref{R-LO}), $\hat{s}$, $\hat{t}$, and
$\hat{u}$ are the parton level Mandelstam variables.  The lowest order
two-to-two matrix element squares in Eq.~(\ref{R-LO}) are given in
Ref.~\cite{Owens:1987mp}. 

The lowest order contributions to the partonic cross sections
$\hat{\sigma}_{ab\rightarrow\gamma_\lambda^*}$ in Eq.~(\ref{Y-hard})
are at order of $O(\alpha_{em}\alpha_s)$ from the
subprocesses: $q\bar{q}\rightarrow \gamma^*(Q)g$ and 
$qg\rightarrow \gamma^*(Q)q$.  Since the subtraction term in
Eq.~(\ref{Y-hard}) starts at order of $O(\alpha_{em}\alpha_s^2)$, 
we have the leading order contributions to the $Y$-term as
\begin{eqnarray}
\frac{d\hat{\sigma}_{ab\rightarrow\gamma^*_\lambda(Q)}^{(Y-LO)}}
     {dQ_T^2\,dy}
&=& 
\frac{d\hat{\sigma}_{ab\rightarrow\gamma^*_\lambda(Q)}^{(LO)}}
     {dQ_T^2\,dy}
\nonumber \\
&=&
\pi\, \frac{\alpha_{em}(\mu)\alpha_s(\mu)}{x_1\,x_2\,S}
\left|
  \frac{1}{eg}\overline{M}_{ab\rightarrow\gamma^*_\mu d}\
  \epsilon^\mu_\lambda(Q) \right|^2 \,
\delta(\hat{s}+\hat{t}+\hat{u}-Q^2)\, ,
\label{Y-LO}
\end{eqnarray}
where $e$ is the electron charge and $\hat{s}$, $\hat{t}$, and
$\hat{u}$ are again the parton level Mandelstam variables.  
The matrix element squares in Eq.~(\ref{Y-LO}) for producing a
polarized virtual photon depend on the photon's polarization vector 
$\epsilon_\lambda^\mu(Q)$.  

The choice of the polarization vectors are not unique \cite{LT-DY}.
But, it should be consistent with the definition of $\alpha_{\rm DY}$
in Eqs.~(\ref{DY-theta}) and (\ref{alpha}).  The angle $\theta$
in Eq.~(\ref{DY-theta}) depends on the frame choice and its $z$-axis
\cite{LT-DY}.  As a result, the extracted value of $\alpha_{\rm DY}$ 
as well as the polarization vectors used to define the polarized
virtual photon cross sections in Eq.~(\ref{alpha}) are also sensitive
to the frame choice and its $z$-axis.  For the consistency of the QCD
factorization formula in Eq.~(\ref{HQT-fac}), it is very important to
use the same polarization vectors to calculate both the virtual
photon fragmentation functions and the $Y$-term.  Since we are 
interested in the region where $Q_T\gg Q$, we will not choose the
Collins-Soper frame \cite{CS-DY}, which offers special advantage when
$Q_T\ll Q$.  Instead, as shown in Fig.~\ref{fig2}, we choose the
$z$-axis along the direction of the virtual photon's momentum
$\vec{Q}$ and angle $\theta$ to be the angle between the $\vec{Q}$ and
$\vec{l}$, the direction of one of the decay leptons.  Our choice of
the $z$-axis corresponds to the ``S-helicity'' frame defined
in Ref.~\cite{LT-DY}.  With our choice of the helicity frame and angle
$\theta$, $\alpha_{\rm DY}=+1(-1)$ corresponds to the virtual photon
in a purely transverse (longitudinal) polarization state.   

To minimize the frame dependence, we present our results in terms of
the covariant hadronic tensors $w_{[ab]\mu\nu}$ and polarization
vector $\epsilon^\mu_\lambda(Q)$, which are defined as
\begin{equation}
\left|
  \frac{1}{eg}\overline{M}_{ab\rightarrow\gamma^*_\mu d}\
  \epsilon^\mu_\lambda(Q) \right|^2 
\equiv w_{[ab]\mu\nu}\, \epsilon^{*\mu}_\lambda(Q)\, 
  \epsilon^\nu_\lambda(Q) \, .
\label{m2-abgd}
\end{equation}
For quark-antiquark annihilation subprocess: 
$q(p_1)\bar{q}(p_2)\rightarrow\gamma^*(Q)g$, we obtain the hadronic
tensor,
\begin{eqnarray}
w_{[q\bar{q}]}^{\mu\nu} 
&=& e_q^2\, \left(\frac{4}{9}\right)
            \left[\frac{1}{\hat{t}\hat{u}}\right]
\Bigg\{
-4Q^2\left[
\left(p_1^\mu-\frac{p_1\cdot Q}{Q^2} Q^\mu\right)
\left(p_1^\nu-\frac{p_1\cdot Q}{Q^2} Q^\nu\right)\right.
\nonumber \\
&\ & {\hskip 1.4in}
+ \left.
\left(p_2^\mu-\frac{p_2\cdot Q}{Q^2} Q^\mu\right)
\left(p_2^\nu-\frac{p_2\cdot Q}{Q^2} Q^\nu\right)\right]
\nonumber \\
&\ & {\hskip 1.0in}
-\left(g^{\mu\nu}-\frac{Q^\mu Q^\nu}{Q^2}\right)
\left[\left(2p_1\cdot Q\right)^2 
     +\left(2p_2\cdot Q\right)^2\right] \Bigg\}
\label{qqb-hard}
\end{eqnarray}
with quark's fractional charge $e_q$; and for the ``Compton''
subprocess: $g(p_1)q(p_2)\rightarrow\gamma^*(Q)q$, we have
\begin{eqnarray}
w_{[gq]}^{\mu\nu} 
&=& e_q^2\, \left(\frac{1}{6}\right)
            \left[\frac{1}{\hat{s}(-\hat{u})}\right]
\Bigg\{
-4Q^2
\left(p_1^\mu-\frac{p_1\cdot Q}{Q^2} Q^\mu\right)
\left(p_1^\nu-\frac{p_1\cdot Q}{Q^2} Q^\nu\right)
\nonumber \\
&\ & {\hskip 1.3in}
- 8Q^2
\left(p_2^\mu-\frac{p_2\cdot Q}{Q^2} Q^\mu\right)
\left(p_2^\nu-\frac{p_2\cdot Q}{Q^2} Q^\nu\right)
\nonumber \\
&\ & {\hskip 1.3in}
- 4Q^2 \left[
\left(p_1^\mu-\frac{p_1\cdot Q}{Q^2} Q^\mu\right)
\left(p_2^\nu-\frac{p_2\cdot Q}{Q^2} Q^\nu\right)\right.
\nonumber \\
&\ & {\hskip 1.7in}
+ \left.
\left(p_2^\mu-\frac{p_2\cdot Q}{Q^2} Q^\mu\right)
\left(p_1^\nu-\frac{p_1\cdot Q}{Q^2} Q^\nu\right)\right]
\nonumber \\
&\ & {\hskip 1.3in}
-\left(g^{\mu\nu}-\frac{Q^\mu Q^\nu}{Q^2}\right)
\left[\left(Q^2-\hat{s}\right)^2 
     +\left(Q^2-\hat{u}\right)^2\right] \Bigg\}\, .
\label{gq-hard}
\end{eqnarray}
With our choice of the helicity frame, we have the polarization vector 
for a longitudinally polarized virtual photon as \cite{LT-DY} 
\begin{equation}
\epsilon^\mu_L(Q) = -\, \frac{Q}{\sqrt{(P\cdot Q)^2 -Q^2 S}}\,
 \left[P^\mu - \frac{P\cdot Q}{Q^2}\, Q^\mu\right]
\label{Ep-L}
\end{equation}
where $P^\mu\equiv P_A^\mu + P_B^\mu$.  The polarization vector
$\epsilon^\mu_L(Q)$ in Eq.~(\ref{Ep-L}) reduces to the unit vector
$\hat{z}^\mu$ in the photon's rest frame \cite{LT-DY}.  By contracting
the hadronic tensors in Eqs.~(\ref{qqb-hard}) and (\ref{gq-hard}) with
the polarization tensor $\epsilon^{*\mu}_L(Q)\epsilon^{\nu}_L(Q)$ (or 
$-g^{\mu\nu}$), one can easily obtain the matrix element squares
for producing longitudinally polarized (or unpolarized) virtual
photons.  Because of the ambiguities in choosing the $x$- or $y$-axis,
the functional forms of the transverse polarization vectors
$\epsilon^{\mu}_{T_i}(Q)$ with $i=1,2$ are not unique.  One can derive
the matrix element squares for producing a transversely polarized
virtual photon by using the polarization tensor 
$\frac{1}{2}\sum_{i=1,2}\epsilon^{*\mu}_{T_i}(Q)\epsilon^{\nu}_{T_i}(Q)$
\cite{LT-DY}, or by taking the one half of the difference
in the matrix element squares for unpolarized and longitudinally
polarized virtual photons. 

%%%%%%%%%%%%%%%%%%%%%%%%%%%%%%%%%%%%%%%%%%%%%%%%%%%%%%%%%%%%%%%%%%%%%

For deriving our numerical results, we use CTEQ5M parton
distributions \cite{Lai:2000wy} and the virtual photon fragmentation
functions from Ref.~\cite{QZ-vpff}.  We set the renormalization,
factorization, and fragmentation scales equal to
$\mu=\kappa \sqrt{Q^2+Q_T^2}$ with $\kappa$ a constant of order one.

With the large $Q_T$ setting up the scale of hard collision, the QCD
factorization formula in Eq.~(\ref{HQT-fac}) should be valid for $Q^2$
as small as a few GeV$^2$.  Since we are interested in the kinematics
similar to J/$\psi$ production at high $Q_T$, we choose $Q=2$ and
5~GeV in following plots.

\begin{figure}
\begin{center}
\begin{minipage}[t]{3in}
\epsfig{figure=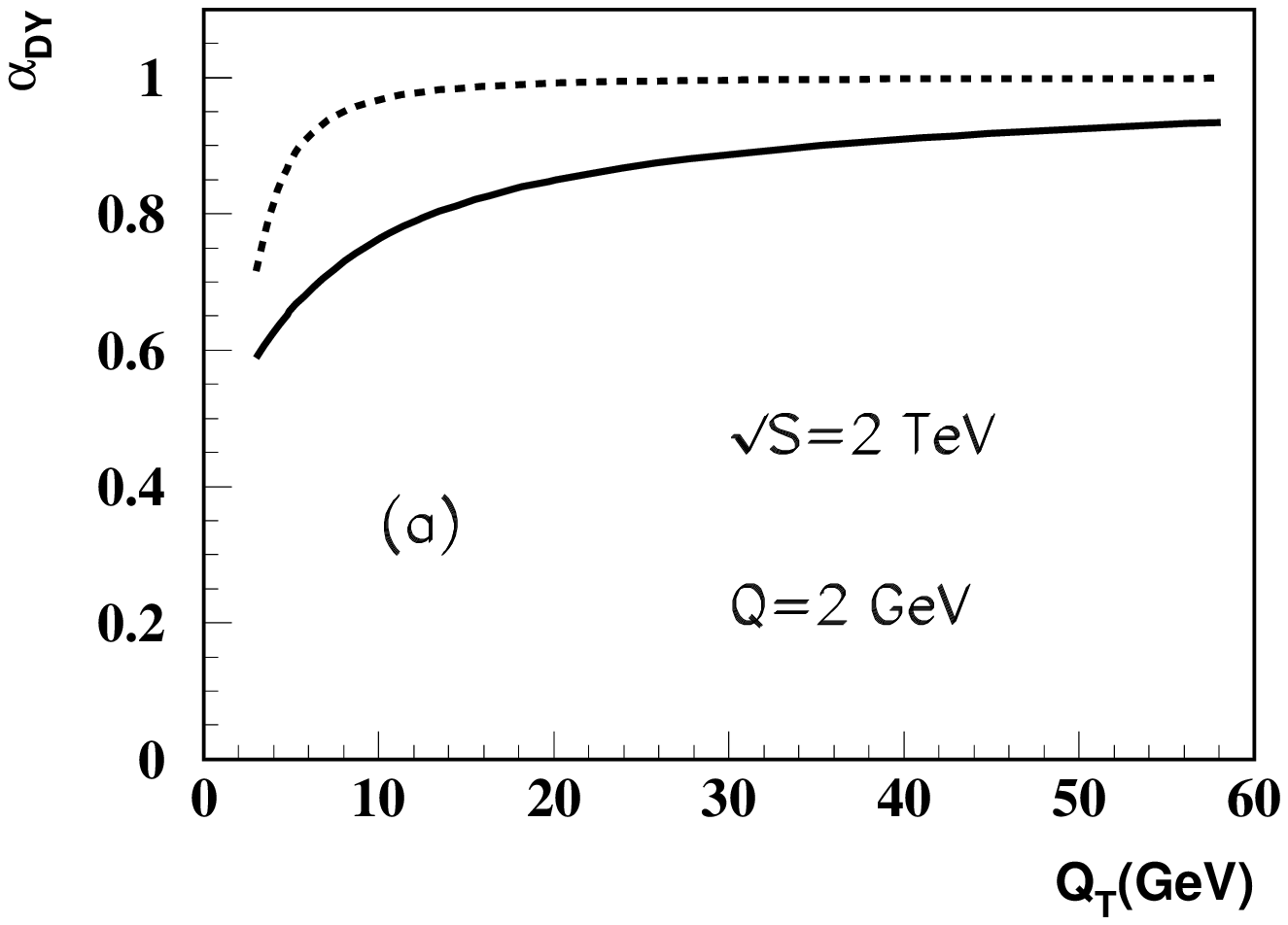,width=3.0in}
\end{minipage}
\hfil
\begin{minipage}[t]{3in}
\epsfig{figure=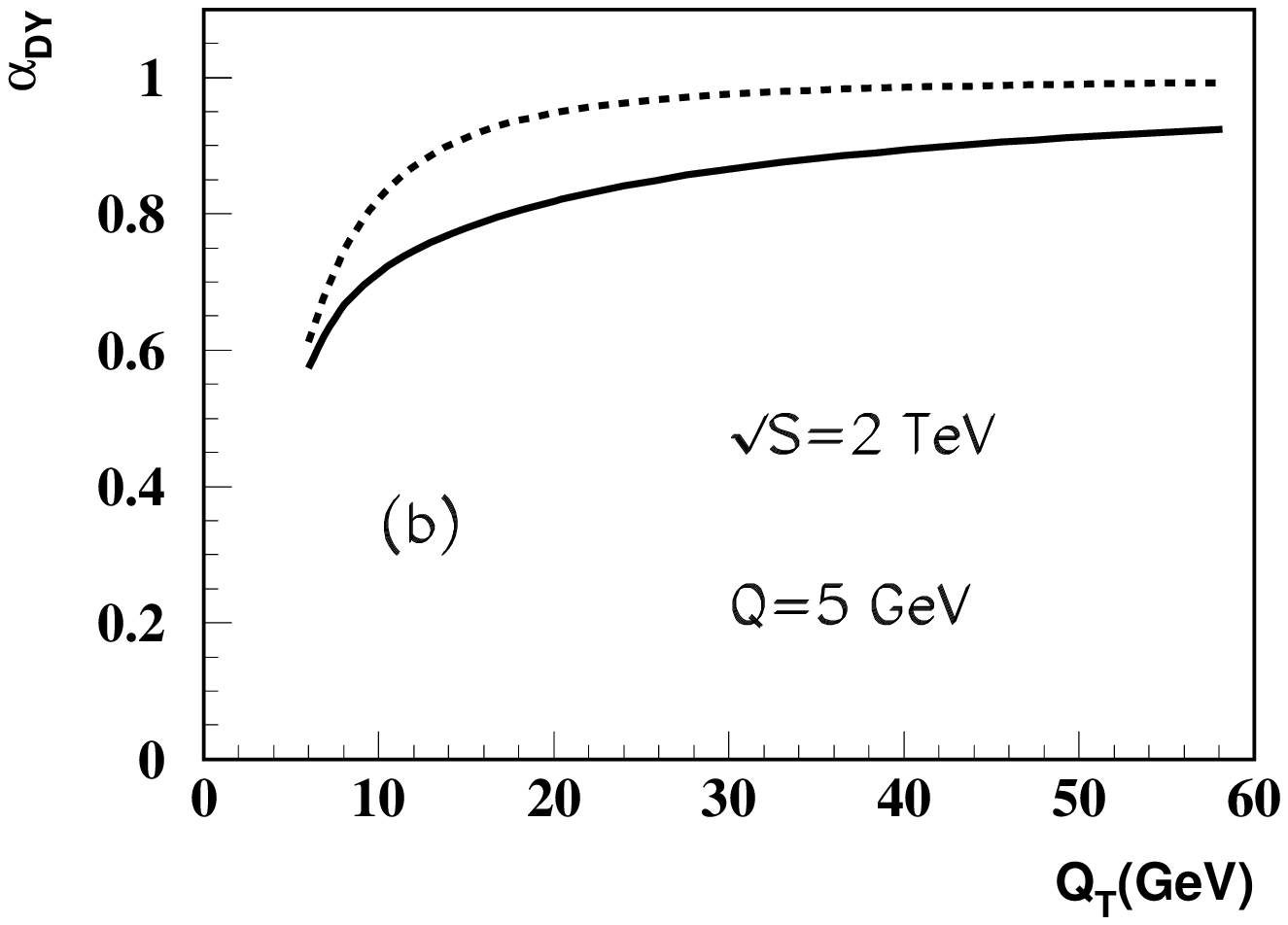,width=3.0in}
\end{minipage}
\end{center}
\caption{$\alpha_{\rm DY}$ as a function of $Q_T$ in proton-antiproton
collisions at $\sqrt{S}=2$~TeV and $Q=2$~GeV (a) and $Q=5$~GeV (b).
Dashed lines correspond to the $Y$-term only, and solid lines include
both the $Y$-term and the resummed contributions. }
\label{fig3}
\end{figure}

\begin{figure}
\begin{center}
\begin{minipage}[t]{3in}
\epsfig{figure=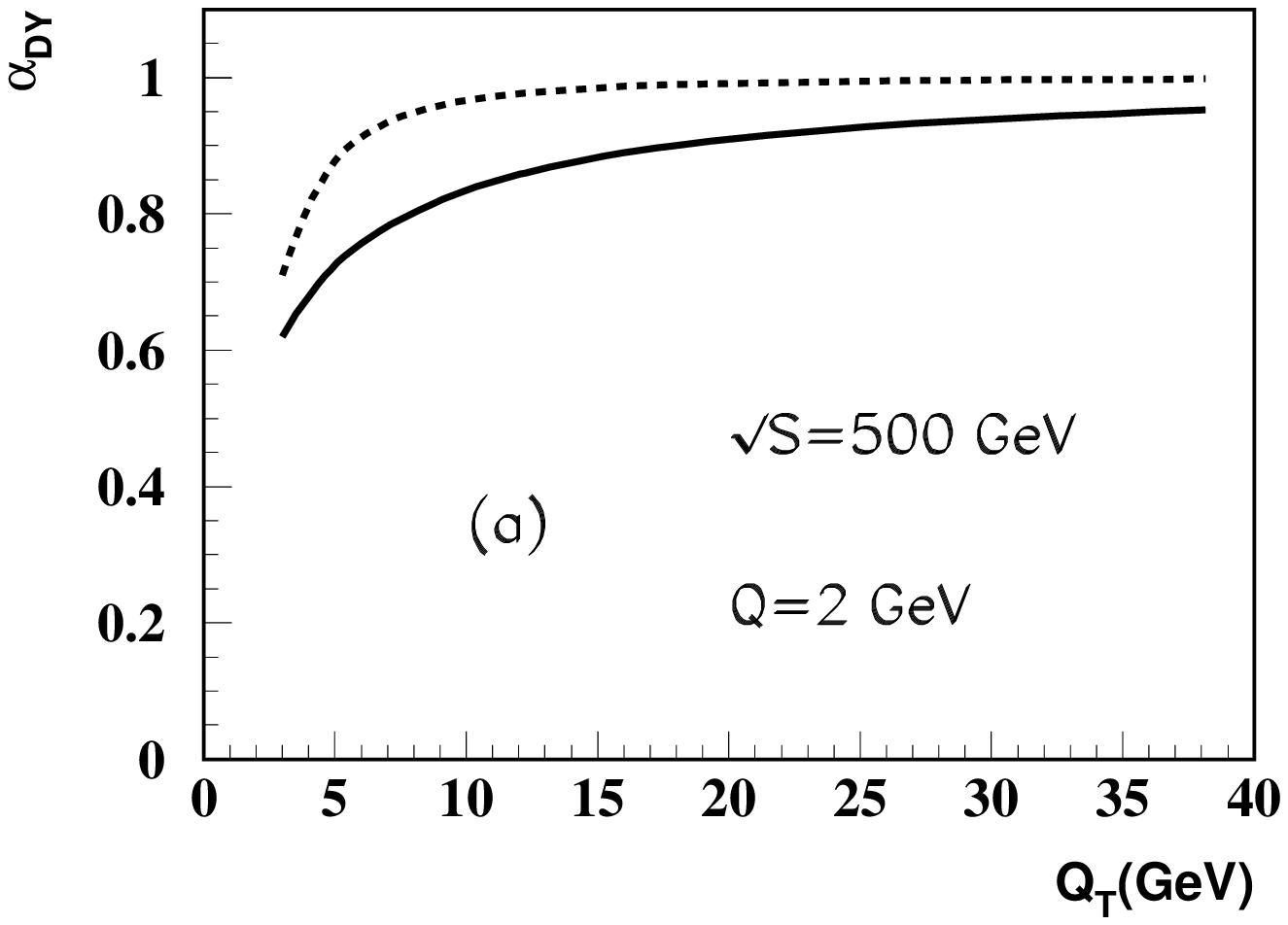,width=3.0in}
\end{minipage}
\hfil
\begin{minipage}[t]{3in}
\epsfig{figure=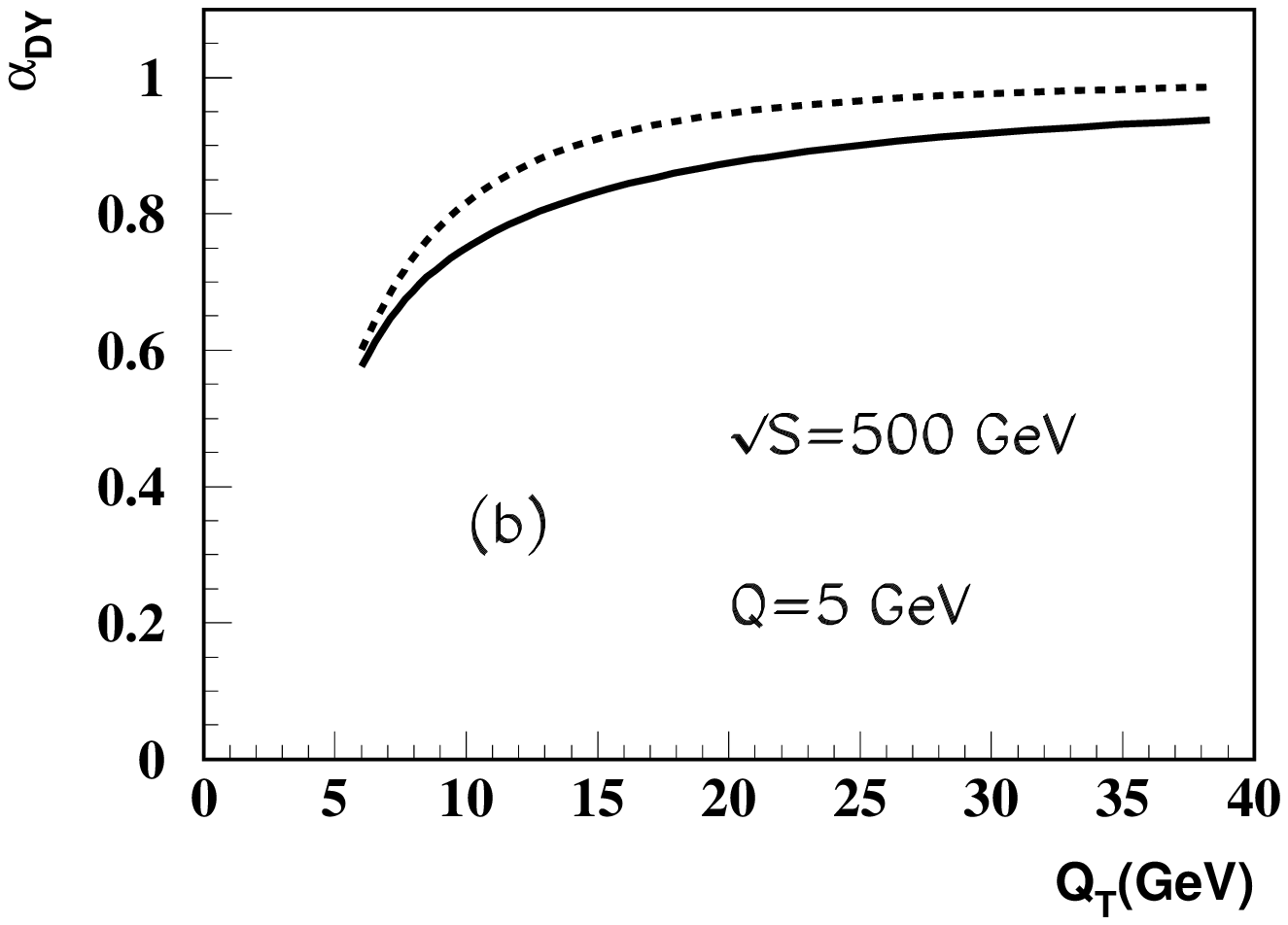,width=3.0in}
\end{minipage}
\end{center}
\caption{$\alpha_{\rm DY}$ as a function of $Q_T$ in proton-proton
collisions at $\sqrt{S}=500$~GeV and $Q=2$~GeV (a) and $Q=5$~GeV (b).
Dashed lines correspond to the $Y$-term only, and solid lines include
both the $Y$-term and the resummed contributions. }
\label{fig4}
\end{figure}

In Fig.~\ref{fig3}, we plot the polarization parameter $\alpha_{\rm
DY}$ in the proton-antiproton collisions as a function of $Q_T$ at
$\sqrt{S}=2$~TeV (new Tevatron energy).  The dashed lines represent 
the $\alpha_{\rm DY}$ with only the leading order $Y$-term derived
from Eqs.~(\ref{qqb-hard}) and (\ref{gq-hard}).  The solid lines
are equal to the $\alpha_{\rm DY}$ with both the $Y$-term and the
resummed contributions.  In Fig.~\ref{fig4}, we plot the same
polarization parameter $\alpha_{\rm DY}$ in the proton-proton
collisions at $\sqrt{S}=500$~GeV (the RHIC energy).  The constant
$\kappa=1$ in both Figs.~\ref{fig3} and \ref{fig4}.  By varying the
$\kappa$ from $1/2$ to $2$, we find that the cross sections are not
very sensitive to the choice of $\kappa$ \cite{BQZ-DY}, and the
numerical values of the $\alpha_{\rm DY}$ are relatively stable.

From Figs.~\ref{fig3} and \ref{fig4}, we see that the virtual
photons in Drell-Yan massive lepton-pair production are likely to be
transversely polarized at high $Q_T$.  The resummed contributions
provide significant corrections to the numerical values of the
polarization parameter $\alpha_{\rm DY}$, and they are more important
at the low $Q^2$.  As pointed out in
Ref.~\cite{QZ-vpff}, the longitudinally polarized virtual photon
fragmentation functions dominate the threshold region.  Consequently,
we expect that the resummed contributions to the cross sections of
longitudinally polarized virtual photons are more important in low
$Q_T$ than high $Q_T$ region, which are clearly evident in the
difference between the solid and dashed lines.  Since the
perturbatively calculable partonic hard parts in the modified QCD
factorization formula in Eq.~(\ref{HQT-fac}) are evaluated at a single
hard scale $Q_T$ without large logarithms, high order corrections
should not alter the polarization parameter $\alpha_{\rm DY}$
dramatically \cite{BQZ-DY}.  We therefore conclude that the virtual
photons of invariant mass $Q$ in Drell-Yan massive lepton-pair
production are likely to be transversely polarized when $Q_T\gg Q$. 

When $Q_T\gg M_{{\rm J/}\psi}$, hadronic J/$\psi$ and the virtual
photon production have a lot in common.  The key difference is the
difference between their respective fragmentation functions, as shown
in Fig.~\ref{fig1}.  As argued in Ref.~\cite{QZ-vpff}, the virtual
photon fragmentation functions are completely perturbative, while the
parton to J/$\psi$ fragmentation functions involve final-state
radiations and soft interactions \cite{QS-jpsi}.  As shown in
Fig.~\ref{fig1}, the 
fragmentation functions from a parton $d$ to a physical J/$\psi$ can
be approximated by the fragmentation functions to a virtual gluon of
invariant mass $Q$, which immediately decay into a $c\bar{c}$-pair,
convoluted with a transition from the $c\bar{c}$-pair to a physical
J/$\psi$ \cite{QS-jpsi}.  The first part of the fragmentation
functions for a parton to a virtual gluon should be perturbatively
calculable.  Since a gluon can directly couple to a gluon, the
evolution equations for the virtual gluon fragmentation functions can 
have a nonvanish leading order inhomogeneous evolution kernel for a
gluon to a virtual gluon of mass $Q$.  Consequently, the
quark-to-virtual-gluon and gluon-to-virtual-gluon fragmentation
functions should be equally important, while the
gluon-to-virtual-photon fragmentation functions are much smaller than
quark-to-virtual-photon fragmentation functions \cite{QZ-vpff}.
However, this difference should not have too much effect on the ratio
of the transverse and longitudinal polarizations.  We then expect that 
at high $Q_T$, the virtual gluon, so as the $c\bar{c}$-pair
immediately produced from the decay of the virtual gluon, are more
likely to be transversely polarized.   

The produced charm and anticharm quark pair of invariant mass $Q$ can
in principle radiate gluons and have soft interactions with other
partons in the collisions.  However, due to the heavy quark mass, such
final-state interactions during the transition from the produced
$c\bar{c}$-pair to a physical J/$\psi$ meson are not expected to
significantly change the polarization \cite{BKL-jpsi}.  If the
formation from the $c\bar{c}$-pair of invariant mass $Q$ to a physical
J/$\psi$ meson does not change the polarization, one can expect the
polarization of the J/$\psi$ mesons produced at high $Q_T$ to be
similar to the polarization of the virtual photon in Drell-Yan massive
lepton-pair production at the same kinematics.  For example, the
non-relativistic QCD (NRQCD) model of J/$\psi$ production precisely
predicts the J/$\psi$ mesons to be transversely polarized at large
$Q_T$ \cite{BKL-jpsi}, which is not consistent with recent  
Fermilab data \cite{CDF-jpsi}.  

Since J/$\psi$ production at high $Q_T$ involves both perturbative and
nonperturbative physics, we have to find answers to the following two
questions in order to resolve this inconsistency.  The first question
is about the reliability of the perturbative calculations for the
relevant kinematics.  The second question is about the separation
between perturbative and nonperturbative physics.  These two questions
can be summarized into one: if there is a reliable QCD factorization
for hadronic J/$\psi$ production \cite{QS-jpsi}.  The measurements of
the virtual photon polarization in Drell-Yan massive lepton-pair
production can help us to answer the first question.  Since the QCD
factorization for Drell-Yan massive lepton-pair production is expected
to be valid when $Q_T\gg Q$ \cite{BQZ-DY}, the measurements of the
polarization parameter $\alpha_{\rm DY}$ provide excellent tests of
the reliability of perturbative calculations for the kinematic region
similar to J/$\psi$ production.  Although the measurements of the
$\alpha_{\rm DY}$ cannot test the final-state effect of J/$\psi$
production or answer the second question, the measured difference
between $\alpha_{\rm DY}$ and $\alpha_{{\rm J/}\psi}$ can help us to
isolate the role of the final-state effect, and to narrow the
questions about the J/$\psi$ production. 

This work was supported in part by the U.S. Department of Energy under
Grant No. DE-FG02-87ER40731.

%%%%%%%%%%%%%% Begin References %%%%%%%%%%%%%%%%%%%%%%%%%%%%%%%%

%%%%%%%%%%%%%% Begin Figure Captions %%%%%%%%%%%%%%%%%%%%%%%%%%%%%%%%%%%

%%%%%%%%%%%%%%%%%%%%%%%%%%%%%%%%%%%%%%%%%%%%%%%%%%%%%%%%%%%%%%%%%%
\end{document}